\documentclass[conference]{IEEEtran}
\IEEEoverridecommandlockouts

\usepackage{cite}
\usepackage{amsmath,amssymb,amsfonts}
\usepackage{algorithmic}
\usepackage{graphicx}
\usepackage{textcomp}
\usepackage{xcolor}

\usepackage{orcidlink}
\usepackage{makecell}
\usepackage{bbding}
	\definecolor{ForestGreen}{RGB}{0, 144, 83}
	\definecolor{Orchid3}{RGB}{205, 105, 201}
	\definecolor{DarkGoldenrod3}{RGB}{206, 150, 11}
\usepackage{siunitx}
\usepackage{listings}
\usepackage{multirow}
\definecolor{LightGray}{gray}{0.7}
\usepackage[T1]{fontenc}  

\def\BibTeX{{\rm B\kern-.05em{\sc i\kern-.025em b}\kern-.08em
    T\kern-.1667em\lower.7ex\hbox{E}\kern-.125emX}}
\begin{document}

\title{MT4G: A Tool for Reliable Auto-Discovery of NVIDIA and AMD GPU Compute and Memory Topologies
\thanks{
This work has been accepted for publication in Workshops of the International Conference for High Performance Computing, Networking, Storage and Analysis (SC Workshops ’25).

This work was supported by the Plasma-PEPSC project, which has received funding from the European High-Performance Computing Joint Undertaking (JU) as well as the German Federal Ministry of Research, Technology, and Space (BMFTR) through Grant Agreement No. 101093261, as well as by the
SEANERGYS project, which has received funding from the European High-Performance Computing Joint Undertaking (JU) under grant agreement No. 101177590.
This work was also supported by BMFTR through the initiative SCALEXA and the PDExa project (16ME0641).
Part of the performance results have been obtained on systems in the test environment BEAST (Bavarian Energy Architecture \& Software Testbed) at the Leibniz Supercomputing Centre.
Grok 3 and Copilot (GPT 4.1) were used to help improve text quality, spelling, and grammar.
}
}

\makeatletter
\newcommand{\linebreakand}{%
  \end{@IEEEauthorhalign}
  \hfill\mbox{}\par
  \mbox{}\hfill\begin{@IEEEauthorhalign}
}
\author{
\IEEEauthorblockN{1\textsuperscript{st} Stepan Vanecek}
\IEEEauthorblockA{\textit{Technical University of Munich} \\
    Garching, Germany \\
    stepan.vanecek@tum.de\,\orcidlink{0009-0008-4120-9472}} 
\and
\IEEEauthorblockN{2\textsuperscript{nd} Manuel Walter Mußbacher}
\IEEEauthorblockA{\textit{Technical University of Munich} \\
    Garching, Germany \\
    manuel.mussbacher@tum.de\,\orcidlink{0009-0006-9776-6994}} 
\and
\IEEEauthorblockN{3\textsuperscript{rd} Dominik Größler}
\IEEEauthorblockA{\textit{Technical University of Munich} \\
    Garching, Germany \\
    dominik.groessler@tum.de\,\orcidlink{0009-0006-7923-7466}} 
\linebreakand
\IEEEauthorblockN{4\textsuperscript{th} Urvij Saroliya}
\IEEEauthorblockA{\textit{Technical University of Munich} \\
    Garching, Germany \\
    urvij.saroliya@tum.de\,\orcidlink{0009-0009-0953-5048}} 
\and
\IEEEauthorblockN{5\textsuperscript{th} Martin Schulz}
\IEEEauthorblockA{\textit{Technical University of Munich} \\
    Garching, Germany \\
    schulzm@in.tum.de\,\orcidlink{0000-0001-9013-435X}} 
}

\maketitle

\begin{abstract}
Understanding GPU topology is essential for performance-related tasks in HPC or AI. Yet, unlike for CPUs with tools like hwloc, GPU information is hard to come by, incomplete, and vendor-specific.

In this work, we address this gap and present MT4G, an open-source and vendor-agnostic tool that automatically discovers GPU compute and memory topologies and configurations, including cache sizes, bandwidths, and physical layouts.
MT4G combines existing APIs with a suite of over 50 microbenchmarks, applying statistical methods, such as the Kolmogorov-Smirnov test, to automatically and reliably identify otherwise programmatically unavailable topological attributes.

We showcase MT4G's universality on ten different GPUs and demonstrate its impact through integration into three workflows: GPU performance modeling, GPUscout bottleneck analysis, and dynamic resource partitioning.
These scenarios highlight MT4G's role in understanding system performance and characteristics across NVIDIA and AMD GPUs, providing an automated, portable solution for modern HPC and AI systems.
\end{abstract}


\section{Motivation and Contributions}

The architectural complexity of modern high-performance chips has increased significantly as a byproduct of the demand for ever-growing computing power. This complexity manifests itself in an increased number of compute and memory/cache resources, connected with varying on-chip network topologies.
The resulting chip architectures typically feature a large number of cores, each with access to core-private L1/L2 caches, while last-level caches (LLCs) and main memory are shared across all compute resources.

Graphics Processing Units (GPUs) increase the topological complexity by providing orders of magnitude more compute resources than CPUs, leading to larger and more complex on-chip topologies.
Additionally, their cache hierarchies are more heterogeneous, comprising specialized units shared among specific resources.
Beyond traditional L1, L2, and occasionally L3 caches, GPU microarchitectures incorporate specialized caches and scratchpad memories, such as Shared Memory on NVIDIA or Local Data Share (LDS) on AMD, directly addressed and accessed with respective load/store instructions.
Some of this complexity is hidden from the user; other aspects, such as Shared Memory or LDS, must be managed directly by the programmer.
Efficient data movement between compute cores and memory elements is critical, as suboptimal memory usage can create significant performance (and power) bottlenecks, particularly for memory-intensive applications.

Optimizing the usage of available memory elements begins with understanding which memory elements are available, what their main attributes are, and where in the topology of the chip they are located.
Various tools and interfaces, like hwloc~\cite{hwloc}, likwid-topology~\cite{treibig2010likwid}, or sys-sage~\cite{vanecek2024sys}, expose the topological information on CPUs (as presented in \texttt{/sys/devices/system}) in an interface suitable both for human and computer processing.
However, no standardized approach to collecting and exposing this information on GPUs is available, significantly complicating the efforts to understand the available hardware.
While interfaces like \texttt{cudaDeviceProp} on the NVIDIA side and \texttt{hipDeviceProp} on both AMD and NVIDIA provide basic information, such as memory size and the number of Streaming Multiprocessors (SMs)/Compute Units (CUs), they lack detailed insights into the cache subsystem or the overall architectural topology. For
more detailed information, developers must read through various vendor-specific datasheets and architectural whitepapers (such as~\cite{amd_cdna3_2024,nvidia_h100_2022}), which still miss essential information, or one-off microarchitecture-specific reverse-engineering studies (such as~\cite{jia2019dissecting,luo2025dissecting,jia2018dissecting}), which contain enough detail but offer no API and are not available for all architectures, especially among AMD GPUs.
Even though these reports are purely manual and microarchitecture-specific, they prove beneficial for the development of our approach, as we adopt multiple proven benchmarks from these papers and use them to verify our results.

To mitigate this, we present \textbf{MT4G} (Memory Topology for GPUs) tool for unified reporting of memory and compute topologies on NVIDIA and AMD GPUs.
MT4G is \textbf{open-source and publicly available at \url{https://github.com/caps-tum/mt4g}}.
It provides detailed information about the topology of compute and memory building blocks of a GPU, including properties such as cache and memory sizes, cache line size, load latency, and accessibility by compute resources.
It gathers information from vendor-specific APIs, where available, but primarily detects configurations based on measurements obtained from running a comprehensive suite of automatically evaluated microbenchmarks.
To ensure reliable results, MT4G implements a variant of the Kolmogorov-Smirnov probability distribution test~\cite{massey1951kolmogorov} to differentiate real pattern changes from outlier measurements, along with a set of other algorithms and heuristics for other data distributions.
\textbf{MT4G works reliably on all recent NVIDIA microarchitectures (from Pascal onward)\footnote{Although on the market at the time of submission, we did not evaluate the Blackwell and Ada Lovelace microarchitectures due to missing access.} and all AMD CDNA GPUs}, delivering both human- and machine-readable output, suitable for developers and automated tools.
As a result, we present MT4G as a vendor-agnostic, all-in-one solution for collecting and reporting GPU compute and memory topology.
Finally, we demonstrate its versatility through three use cases, in which MT4G provides critical topological information otherwise unavailable from one source or API.
The demonstration scenarios provide critical support for tasks in GPU performance modeling, GPU performance analysis and tuning, and understanding GPU dynamic configuration behavior.

\section{Background}

In this work we focus on detecting the topologies of GPUs in modern HPC systems. In particular, recent NVIDIA (from Pascal onward) and the AMD CDNA microarchitectures.
We do \textbf{not} consider AMD's RDNA GPUs due to their focus on graphics rather than on HPC/AI workloads.
Given AMD's plans to unify RDNA and CDNA into UDNA~\cite{amd_udna_unification_2024}, we will extend the support to UDNA GPUs in the future, but this is beyond the scope of this paper.

\subsection{GPU Architecture}
The success of GPUs in HPC and AI in the last years stems from their massive parallelism, with modern GPUs, like NVIDIA's B100 and AMD's MI300X, featuring tens of thousands of CUDA cores/Stream Processors, far surpassing the tens to hundreds cores on CPUs\footnote{CPU cores often include vector units for parallel operations, akin to GPUs.}.
These compute units are organized hierarchically: groups of 32 NVIDIA or 64 AMD cores form one warp/wavefront\footnote{If two terms are mentioned, like here, the syntax in our paper is <NVIDIA term>/<AMD term>. If one term is present, we follow the NVIDIA terminology, but it may mean the general concept applicable to both vendors.}. It executes instructions in lockstep.
Warps/wavefronts are grouped into Streaming Multiprocessors (SM)/Compute Units(CU), which 
form higher-level clusters like Graphics Processing Clusters (GPCs, NVIDIA) or Accelerator Complex Dies (XCDs, AMD)~\cite{nvidia_h100_2022,amd_cdna3_2024}.
The rich and complex topology is in place to facilitate management of these resources -- from scheduling to providing logical and physical contexts for communication and synchronization. 

The memory hierarchy is equally complex -- comprising L1, L2 (sometimes even L3) caches, with the low-level ones serving a single SM/CU, and higher-level caches shared on the GPU level.
Moreover, there are NVIDIA's Shared Memory or AMD's Local Data Store(LDS) areas, serving as scratchpad memory, along with specialized caches, such as Texture or Readonly cache.~\cite{jia2019dissecting}
Understanding and using this complex memory hierarchy is critical to avoid performance bottlenecks in GPU kernels.

\subsection{HIP, PTX, and AMDGCN}

To characterize GPU topologies, MT4G employs GPU kernels (functions executed on the target GPU).
We employ HIP (Heterogeneous-computing Interface for Portability) from AMD's ROCm stack to write portable kernels -- a single source code compiled with \texttt{nvcc} for NVIDIA GPUs or with LLVM for AMD GPUs~\cite{amd_hip_develop_index}.
\texttt{nvcc} generates an intermediate representation, called Parallel Thread Execution (PTX) ISA -- a low-level virtual assembly with an unlimited register count~\cite{loop-unroll}.
Similarly, LLVM generates AMDGCN ISA for AMD GPUs.
While not vendor-agnostic, PTX and AMDGCN inline assembly can be used directly in the kernel for precise instruction control and to avoid potential compiler optimizations.

\subsection{Change Point Detection (CPD)}

Change Point Detection (CPD) is commonly used to identify structural changes in a series of values at certain points of interest (change points) -- changes in the mean, the variance, or the distribution itself.
It is widely used in fields like signal and speech processing~\cite{harchaoui2009regularized}, 
finance~\cite{lavielle2007adaptive,bai1998estimating}, network analysis and security~\cite{levy2009detection,lung2012distributed}, or bioinformatics~\cite{frick2014multiscale}.
We use CPD in this work to automatically retrieve topological information.
Depending on the application, CPD is classified into online (real-time analysis) and offline (post-mortem evaluation) detection techniques.
Offline algorithms are further divided into parametric (require input parameters, e.g., PELT~\cite{killick2012optimal}, CUSUM~\cite{tartakovsky2014sequential}), and non-parametric ones (e.g., K-S test~\cite{massey1951kolmogorov}, Cremer-von Mises~\cite{48d07914-60c5-30cf-a135-f9a3c16e3a01}, or TopRank~\cite{levy2009detection}).~\cite{truong2020selective,aminikhanghahi2017survey}

We define a series of values $S = {x_1, x_2, ..., x_n}$, where $x_i$ represents a (one- or multi-dimensional) observation result.
On the high level, the problem can be presented as a search for the best segmentation $\tau$ according to the criterion $V(\tau,S)$, which is a sum of costs of all the segments in $\tau$.
The cost function measures how well the data of the sub-signal, as segmented by $\tau$, fits together in terms of its distribution. Thus, minimizing the criterion $V(\tau,S)$ leads to the best possible segmentation of the value series.
Different CPD algorithms use different cost functions, which can be interpreted as a measure of homogeneity of the observations $x_i$ within each segment~\cite{grossler2022capturing,truong2020selective}.

\subsubsection{The Kolmogorov-Smirnov (K-S) test}

The CPD is applied after the benchmark run. Hence, we can rely on offline algorithms.
Given our vendor-agnostic focus and broad range of different benchmarks, we omit making assumptions about the data, and thus we favor non-parametric CPD models~\cite{truong2020selective}.
Selecting the correct statistical tests for CPD depends on the dimensionality as well as the type of change that should be detected.
For multidimensional data, multiple options exist to handle CPD, like testing the Maximum Mean Discrepancy MMD~\cite{rabanser2019failingloudlyempiricalstudy}, using Multirank statistics~\cite{LungYutFong2011RobustCD} or with support vectors~\cite{Camci2010ChangePointDetectionUsingSupportVectors}.
Another option is to reduce the multidimensional problem into a one-dimensional one by applying a reduction function, which also reduces the mathematical complexity, and thus with that the runtime. 
Once reduced, one-dimensional data can be tested for a change point using a two-sample statistical test comparing the distribution on the lower side to the one on the higher side of the alleged change point.
Specifically, among CPDs that deliver one single change point incorporating some quality or confidence metric, we opt for the K-S test to evaluate our raw benchmark results.

The K-S test is a non-parametric statistical test used to compare two independent groups of values following two distribution functions $F(X)$ and $G(X)$~\cite{Wilcox2021IntroductionToRobustEstimationAndHypothesisTesting} using the hypothesis $H_0 : F(X) = G(X) $.
The test calculates an estimation of the Kolmogorov distance value $D = max_x |(F(x) - G(x))|$ as a test statistic that is compared against a critical value to confirm or deny the null hypothesis $H_0$.
According to~\cite{Wilcox2021IntroductionToRobustEstimationAndHypothesisTesting}, we can approximate the critical value by using
\begin{equation}
    d_\alpha = \sqrt{\frac{1}{2} \cdot \frac{n+m}{n\cdot m} \cdot \log (\alpha / 2)}
\end{equation}

\subsection{Related Work} \label{sec:related}

Unlike CPU topology tools, such as hwloc~\cite{hwloc} or likwid~\cite{treibig2010likwid}, no tool can provide comprehensive and vendor-agnostic topological information on GPUs\footnote{We extend sys-sage~\cite{vanecek2023syssage} to GPU topologies through MT4G integration (cf.~Sec.~\ref{sec:sys-sage}).}.
Vendor-specific tools and APIs, such as NVIDIA's {\small\texttt{nvml}}~\cite{nvidia2025nvml}, {\small\texttt{cudaDeviceProp}}, and NSight Compute~\cite{nvidia2025nsightcompute}, and AMD's rocm-smi~\cite{amd2025rocmsmi}, \texttt{hipDeviceProp}, or rocprof~\cite{amd2025rocprof}, offer partial information, but lack the needed detail, especially on the cache hierarchy.
MT4G integrates these interfaces wherever possible to avoid unnecessary benchmarking of information available elsewhere.

Reverse-engineering studies on specific microarchitectures, such as for NVIDIA's Volta~\cite{jia2018dissecting}, Turing~\cite{jia2019dissecting}, or Hopper~\cite{luo2025dissecting} are available, and especially the former studies by Jia et al. provide valuable information with a great level of detail.
We use them in Section~\ref{sec:validation} to validate our results, where no official sources are available.
However, they are one-off microarchitecture-specific studies, which are not available as an automated tool, nor are they vendor-agnostic.

\section{Scope of Provided Information} \label{sec:scope}

Before detailing our tooling approach MT4G, we discuss the scope of information targeted and reported by our tool. In particular, 
there are three main areas of the reported information:
\begin{enumerate}
	\item General information (e.g., vendor, clock size)
	\item Compute resource information (e.g., cores, SMs/CUs)
	\item Memory resource information (e.g., cache size, latency)
\end{enumerate}

While NVIDIA and AMD differ in their architecture, topologies, and the information available via various interfaces, MT4G unifies the output to provide a consistent, vendor-agnostic report of the topology analysis.
The following subsections introduce the details.

\subsection{General Information}\label{sec:general-information}
The general information is retrieved from vendor APIs, namely from HIP and its structure \texttt{hipDeviceProp\_t}~\cite{amd2025hipdeviceprop}, which mimics the NVIDIA structure \texttt{cudaDeviceProp}~\cite{nvidia2025cudadeviceprop}.
It includes: 
\begin{itemize}
	\item GPU vendor, model
	\item GPU clock rate
	\item Compute capability
\end{itemize}

\subsection{Compute Resource Information}\label{sec:compute-resource-information}

Compute resource information provides the number of parallel compute resources and their hierarchical grouping and connectivity, which forms the backbone of the respective topology of the GPU compute resources. Once obtained it then also enables our tool to map the matching memory resources.

\begin{itemize}
    \item Number of SMs/CUs
    \item Max. number of blocks per SM/CU
    \item Max. number of threads per block and SM/CU
	\item Number of cores and warps/SIMD per SM/CU
    \item Warp size
    \item Number of registers per block and SM/CU\footnote{Although de-facto memory-related, registers are mentioned here due to their close ties to compute resources and distinct structure of the provided information.}
    \item Mapping of logical to physical CU ids (AMD only)
\end{itemize}

The number of cores per SM/CU comes from a microarchitecture-specific internal lookup table; the rest from \texttt{hipDeviceProp\_t}.

The majority of information from Sections~\ref{sec:general-information} and~\ref{sec:compute-resource-information} is already available through different APIs and tools.
MT4G includes this information for the sake of completeness -- providing comprehensive information on the entire GPU topology via a single interfaces -- complementing the main focus on memory subsystems.

\subsection{Memory Resource Information}

\begin{table*}[t]
    \centering
    \begin{tabular}{| c || c | c | c | c | c | c | c |}
    \hline
    \textbf{Memory Element} & \textbf{Size} & \textbf{\makecell{Load \\ Latency}} & \textbf{\makecell{Read \& Write \\ Bandwidth}} & \textbf{\makecell{Cache \\ Line Size}} & \textbf{\makecell{Fetch \\ Granularity}} & \textbf{\makecell{Amount per \\ SM/CU or GPU}} & \textbf{\makecell{Physically \\ Shared With}} \\
    \hline
    \hline
    \multicolumn{8}{|c|}{\raisebox{-0.5ex}{\LARGE \textbf{NVIDIA}}} \\
    \hline
    \textbf{L1 cache} & \textcolor{ForestGreen}{\Checkmark} & \textcolor{ForestGreen}{\Checkmark} & \textdagger & \textcolor{ForestGreen}{\Checkmark} & \textcolor{ForestGreen}{\Checkmark} & \textcolor{ForestGreen}{\Checkmark} & \textcolor{ForestGreen}{\Checkmark} {\small (RO,TEX,CO) } \\
    \textbf{L2 cache} & \textcolor{ForestGreen}{\Checkmark}{\small (API)} & \textcolor{ForestGreen}{\Checkmark} & \textcolor{ForestGreen}{\Checkmark} & \textcolor{ForestGreen}{\Checkmark} & \textcolor{ForestGreen}{\Checkmark} & \textcolor{ForestGreen}{\Checkmark} & {\small \textit{n/a}} \\
    \textbf{Texture cache} & \textcolor{ForestGreen}{\Checkmark} & \textcolor{ForestGreen}{\Checkmark} & \textdagger & \textcolor{ForestGreen}{\Checkmark} & \textcolor{ForestGreen}{\Checkmark} & \textcolor{ForestGreen}{\Checkmark} & \textcolor{ForestGreen}{\Checkmark} {\small (L1,RO,CO) } \\
    \textbf{Readonly cache} & \textcolor{ForestGreen}{\Checkmark} & \textcolor{ForestGreen}{\Checkmark} & \textdagger & \textcolor{ForestGreen}{\Checkmark} & \textcolor{ForestGreen}{\Checkmark} & \textcolor{ForestGreen}{\Checkmark} & \textcolor{ForestGreen}{\Checkmark} {\small (L1,TEX,CO) } \\
    \textbf{Constant L1 cache} & \textcolor{ForestGreen}{\Checkmark} & \textcolor{ForestGreen}{\Checkmark} & \textdagger & \textcolor{ForestGreen}{\Checkmark} & \textcolor{ForestGreen}{\Checkmark} & \textcolor{ForestGreen}{\Checkmark} & \textcolor{ForestGreen}{\Checkmark} {\small (L1,RO,TEX) } \\
    \textbf{Constant L1.5 cache} & \textcolor{ForestGreen}{\Checkmark} {\small (\SI{64}{\kibi\byte}) } & \textcolor{ForestGreen}{\Checkmark} & \textdagger & \textcolor{ForestGreen}{\Checkmark} & \textcolor{ForestGreen}{\Checkmark} & {\small\XSolid} & {\small \textit{n/a}} \\
    \textbf{Shared Memory} & \textcolor{ForestGreen}{\Checkmark} {\small (API)} & \textcolor{ForestGreen}{\Checkmark} & \textdagger & {\small \textit{n/a}} & {\small \textit{n/a}} & {\small \textit{n/a}} & {\small \textit{n/a}} \\
    \textbf{Device Memory} & \textcolor{ForestGreen}{\Checkmark} {\small (API)} & \textcolor{ForestGreen}{\Checkmark} & \textcolor{ForestGreen}{\Checkmark} & {\small \textit{n/a}} & {\small \textit{n/a}} & {\small \textit{n/a}} & {\small \textit{n/a}} \\

    \hline
    \hline
    
    \multicolumn{8}{|c|}{\raisebox{-0.5ex}{\LARGE \textbf{AMD}}} \\
    \hline
    \textbf{vL1 cache} & \textcolor{ForestGreen}{\Checkmark} & \textcolor{ForestGreen}{\Checkmark} & \textdagger & \textcolor{ForestGreen}{\Checkmark} & \textcolor{ForestGreen}{\Checkmark} & \textcolor{ForestGreen}{\Checkmark} & {\small \textit{n/a}} \\
    \textbf{sL1d cache} & \textcolor{ForestGreen}{\Checkmark} & \textcolor{ForestGreen}{\Checkmark} & \textdagger & \textcolor{ForestGreen}{\Checkmark} & \textcolor{ForestGreen}{\Checkmark} & {\small \textit{n/a}} & \textcolor{ForestGreen}{\Checkmark} {\small (CU id)}  \\
    \textbf{L2 cache} & \textcolor{ForestGreen}{\Checkmark} {\small (API)} & \textcolor{ForestGreen}{\Checkmark} & \textcolor{ForestGreen}{\Checkmark} & \textcolor{ForestGreen}{\Checkmark} {\small (API)} & \textcolor{ForestGreen}{\Checkmark} &  \makecell{\textcolor{ForestGreen}{\Checkmark} {\small (API)}} & {\small \textit{n/a}}   \\
    \textbf{L3 cache} & \textcolor{ForestGreen}{\Checkmark} {\small (API)} & {\small\XSolid} & \textcolor{ForestGreen}{\Checkmark} & \textcolor{ForestGreen}{\Checkmark}{\small (API)} & {\small\XSolid} & \textcolor{ForestGreen}{\Checkmark} {\small (API)} & {\small \textit{n/a}} \\
    \textbf{LDS} & \textcolor{ForestGreen}{\Checkmark} {\small (API)} & \textcolor{ForestGreen}{\Checkmark} & \textdagger & {\small \textit{n/a}} & {\small \textit{n/a}} & {\small \textit{n/a}} & {\small \textit{n/a}} \\
    \textbf{Device Memory} & \textcolor{ForestGreen}{\Checkmark} {\small (API)} & \textcolor{ForestGreen}{\Checkmark} & \textcolor{ForestGreen}{\Checkmark} & {\small \textit{n/a}} & {\small \textit{n/a}} & {\small \textit{n/a}} & {\small \textit{n/a}} \\

    \hline
    \end{tabular}

    \raggedright{\textcolor{ForestGreen}{\Checkmark} = Available ; \textcolor{ForestGreen}{\Checkmark}{\small (API)} = Available via an interface {\small (not benchmarked)} ; {\small \XSolid} = Not Available ; {\small \textit{n/a}} = Not applicable \\ \textdagger = We only measure bandwidth on higher-level caches or main memory.}
    \caption{Coverage of provided information and attributes on different memory elements.}
    \label{tab:benchmarks}
\end{table*}

The primary contribution of MT4G is the detailed analysis of the memory subsystem of the GPU.
The information is often not accessible programmatically --- neither through vendor-specific (such as nvidia-smi) nor vendor-agnostic or third-party tools (such as \texttt{likwid-topology}).
Some information is even unavailable in the publicly available documentation or whitepapers.
By integrating API data with tailored microbenchmarks, MT4G is the first automated tool to report attributes like cache sizes, latencies, and bandwidths, critical for optimizing HPC and AI workloads on GPUs. 
It covers L1, L2, Readonly, Texture, Constant L1 and L1.5 caches and Shared and Device Memory for NVIDIA GPUs, and sL1d, vL1\footnote{Scalar and Vector L1 Data Caches}, L2, L3 caches, LDS, and Device Memory on AMD GPUs.
Additionally, memory clock rate and bus bandwidth data is also retrieved.

Table~\ref{tab:benchmarks} presents a detailed coverage of the provided information.
Where available, MT4G obtains the information from APIs and drivers: \texttt{hipDeviceProp\_t} for L2 cache, LDS, and Shared and Device memory sizes, HSA runtime library~\cite{hsafoundation2021runtime} to get all cache sizes on AMD GPUs, and KFD kernel driver files~\cite{linuxkernel2025amdgpu} to retrieve AMD GPUs' cache line sizes.
The remaining information (unless marked as coming from an API), is reverse-engineered through microbenchmarks, further described in Section~\ref{sec:benchmarks}.
The reported information includes (unless marked as not applicable or available):
\begin{itemize}
    \item Size: Capacity of the cache or memory element. (obtained via API or microbenchmarks)
    \item Load Latency: Measured via pointer-chase benchmarks targeting specific memory elements.
    \item Read \& Write Bandwidth: Measured via stream-like pattern with vector instructions executed in parallel. Only higher-level caches and Device Memory are benchmarked.
    \item Cache Line Size: Size of the contiguous data segment stored in the cache. (obtained via API or microbenchmarks)
    \item Fetch Granularity: Size of one fetch transaction to the cache (may be different than Cache Line Size, cf. Sec.~\ref{sec:fetch-granularity}).
    \item Amount per SM/CU or GPU: There may be more than one cache per SM/CU or GPU; MT4G retrieves the number of distinct independent memory elements.
    \item Physically Shared With: On NVIDIA: does each logical memory space cache (e.g., Texture, Readonly cache) have its own physical cache, or is it shared with other logical memory spaces?
    On AMD: CU ids of CUs that share one sL1d cache.
\end{itemize}

While MT4G covers a broad range of features and topology details, a few items can currently not be covered: 
MT4G currently is not able to obtain load latency and fetch granularity information on AMD L3 caches (newly introduced in CDNA3), which we plan to add after conducting more experiments on this architecture.
Second, due to the \SI{64}{\kibi\byte} constant array limitation on NVIDIA (cf.~\cite{nvidia2025ptx}), the CL1.5 amount cannot be properly evaluated either.

\section{Benchmarks and Autoevaluation} \label{sec:benchmarks}

The primary contribution of MT4G lies in its use of tailored microbenchmarks to discover the characteristics of the memory subsystem of the GPU not available via official API calls, enabling detailed topology analysis for HPC and AI workloads.
These benchmarks produce specific memory access patterns on the target memory element to expose performance cliffs caused by memory structures, and then rely on K-S tests and other methods to automatically evaluate the results.

Unless noted otherwise, the benchmarks are run sequentially (1 block with 1 thread).
The setup, configuration, post-processing, and evaluation steps are executed on the CPU, while the actual benchmarking is performed on the GPU.
MT4G assumes exclusive usage of the SM/CU or the GPU to ensure reliable results, as noise from concurrent tasks might distort the benchmarking process.

\subsection{Pointer-chase} \label{sec:p-chase}
The pointer-chase (p-chase) algorithm underpins most of our microbenchmarks.
It executes a sequence of load instructions, each load dependent on the previous load's result, ensuring controlled sequential execution.
We adopt the fine-grained p-chase microbenchmark from~\cite{mei2016dissecting}, which records the time (latency) of each individual load instruction, enabling more distinct analysis of the latency statistical distribution, compared to aggregates, such as average.
One p-chase step is shown in Listings~\ref{lst:p-chase-ptx} and ~\ref{lst:p-chase-amdgcn} for NVIDIA and AMD L1 caches, respectively.
Each follows the pattern: (1) start the clock, (2) load one piece of data, (3) stop the clock.
\footnote{The overhead of the clock measurement (and other instructions) is included in the reported load time, however, the constant overhead affects neither the K-S test, nor the tendencies in the result values.}
One run through the entire p-chase array generates a latency for each load, however, for efficiency reasons, we only store the first N results, which is enough to capture repeating patterns.

\lstset{ %
language={[x86masm]Assembler},                
basicstyle=\footnotesize,       
numberstyle=\footnotesize,      
stepnumber=1,                   
numbersep=5pt,                  
backgroundcolor=\color{white},  
showspaces=false,               
showstringspaces=false,         
showtabs=false,                 
frame=single,           
tabsize=2,          
captionpos=b,           
breaklines=true,        
breakatwhitespace=false,    
escapeinside={\%*}{*)}          
}
\begin{figure}[b]
\begin{lstlisting}[label={lst:p-chase-ptx},caption={Time measurement PTX inline assembly for NVIDIA.}]
mov.u32 %0, %%clock;\n\t         // start = clock()
ld.global.ca.u32 %1, [%3];\n\t   // index = *addr
st.shared.u32 [smem_ptr64], %1;  // shared mem store
mov.u32 %2, %%clock;\n\t         // end = clock()
\end{lstlisting}
\end{figure}

\begin{figure}[b]
\begin{lstlisting}[label={lst:p-chase-amdgcn},caption={Time measurement AMDGCN inline assembly for AMD GPUs.}]
s_waitcnt lgkmcnt(0)\n\t         // scalar mem fence
s_waitcnt vmcnt(0)\n\t           // vector mem fence
s_memtime %0\n\t"                // start = clock()
flat_load_dword %1, %3\n\t       // index = *addr;
s_waitcnt lgkmcnt(0)\n\t         // scalar mem fence
s_waitcnt vmcnt(0)\n\t           // vector mem fence
s_memtime %2\n\t                 // end = clock();
\end{lstlisting}
\end{figure}

Before measuring the load time, MT4G performs a warm-up phase (initial pass through the entire array without time measurement).
This step ensures that the array is loaded to the benchmarked memory element, to enable valid measurements.

\begin{figure}[t]
    \centering
    \includegraphics[width=1.02\linewidth]{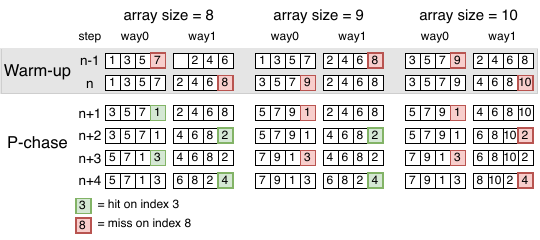}
    \caption{Pointer-chase with different array sizes on a simplified 2-way cache depiction. If the array fits into the cache, the load causes a cache hit; otherwise, a cache miss occurs. Around the boundary, we may experience both hits and misses, as is the case in the middle example.}
    \label{fig:p-chase}
\end{figure}

\subsection{Size benchmarks} \label{sec:size-benchmarks}
The size benchmarks are fundamental to MT4G, as several others were conceptually derived from it.
It is based on the execution of the p-chase benchmark with different array sizes and observing the latencies
If the array is larger than the tested memory element, the latency rises significantly (cache miss) .
The idea is shown in Fig.~\ref{fig:p-chase}: an array exceeding the cache size is (partially) evicted before being accessed in the timed round, resulting in a cache miss.
The K-S test analyzes the latency measurements to identify statistically significant changes in the distribution, pinpointing the cache size of the measured element.

\subsubsection{Workflow}

To balance the versatility and reliability with execution time, we propose the following workflow:
\begin{enumerate}
    \item Identify a narrower search interval
    \item Run the p-chase benchmarks with varying array sizes
    \item Check for outliers; widen the interval and repeat (2) if found
    \item Detect the change point using the K-S test
\end{enumerate}

\paragraph*{(1)} MT4G does not make narrower assumptions about the cache size and starts with a wide search space of \SIrange{1}{1024}{\kibi\byte}.
However, there may be multiple change points -- cache size boundaries, such as L1 and L2 caches -- in this space. 
Therefore, MT4G starts at the lower boundary and exponentially doubles the p-chase size until the array size exceeds the cache size.
Then, binary search narrows the interval again to reduce measurement overhead.

\paragraph*{(2)} The P-chase benchmark is executed with array sizes ranging from the lower to the upper bound of the interval.
The step size is the Fetch Granularity (minimal unit of data in a load operation; cf. Section~\ref{sec:fetch-granularity}): a lower value would inevitably lead to redundant accesses to the same sector and a larger step size could lead to skipping entire cache lines, skewing the measurements.

\paragraph*{(3)} The results are checked for outliers, especially ones caused by cache sizes close to one of the boundaries or unexpected disturbances.
If outliers are found, the search interval is widened, and step (2) is repeated.

\paragraph*{(4). The K-S test}

\begin{figure}[t]
\hspace{-5pt}
\includegraphics[width=1.02\linewidth]{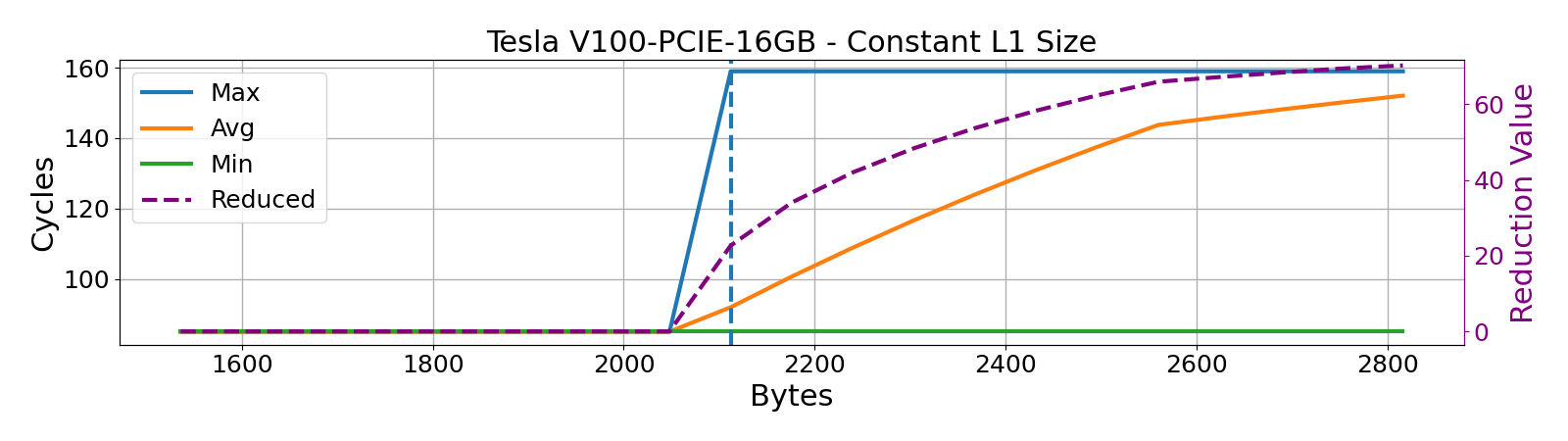}
\vspace{-10pt}
\hspace{-5pt}
\includegraphics[width=1.02\linewidth]{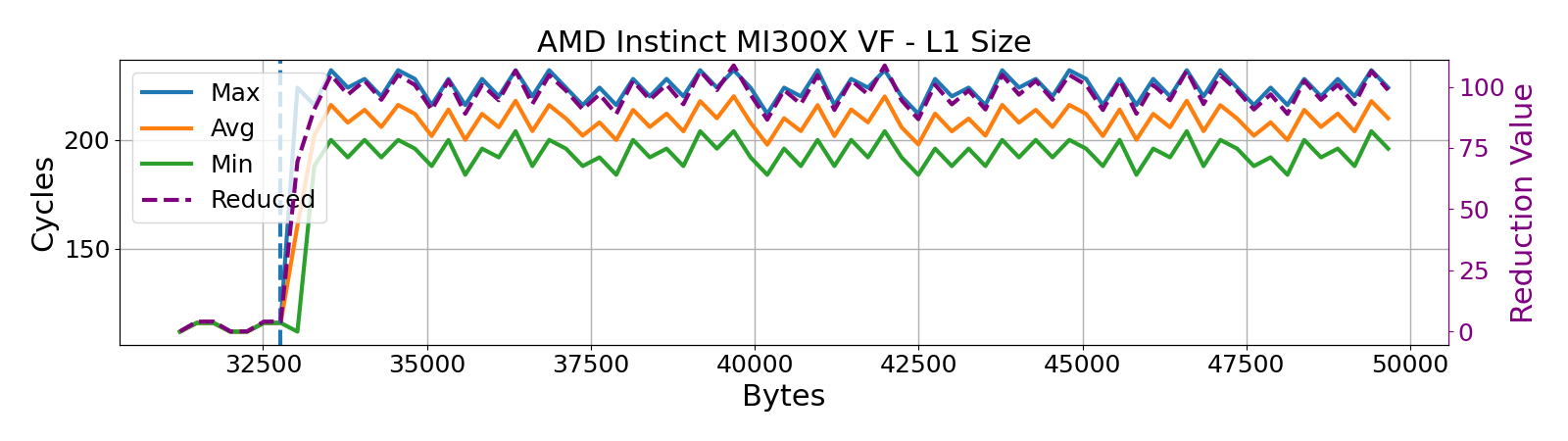}
\vspace{-10pt}
\hspace{-5pt}
\includegraphics[width=1.02\linewidth]{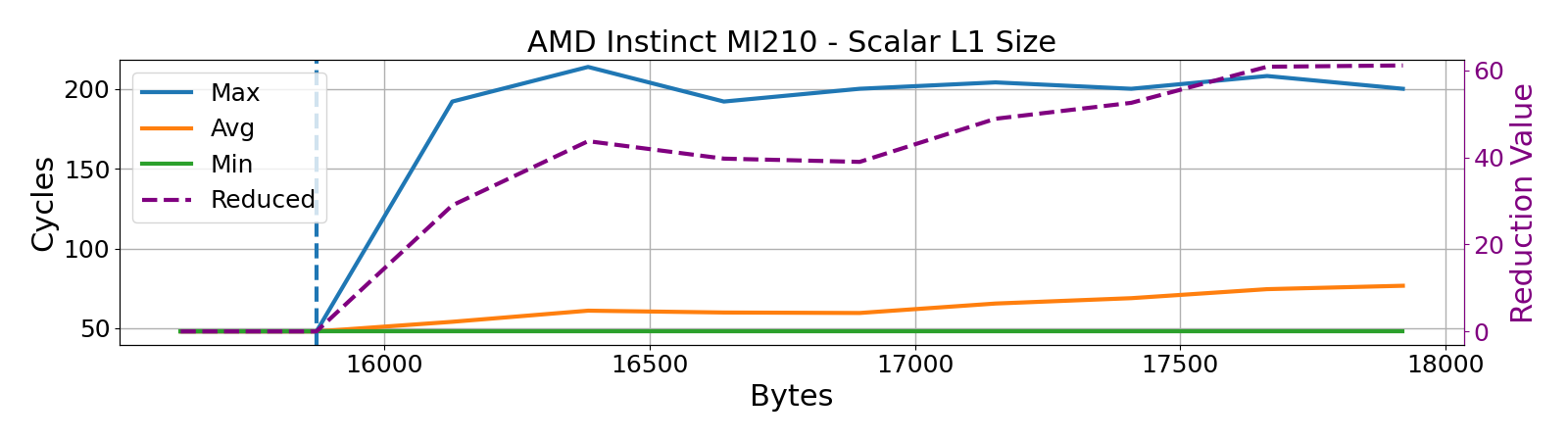}
\caption{NVIDIA V100 CL1, AMD MI300 vL1 and AMD MI210 sL1d example sizes showing raw data analysis (blue,orange,green) and reduction value(violet). Reduction presents the change point (vertical dashed line) most clearly (maximum is prone to outliers).}
    \label{fig:raw-data}
\end{figure}

The runs of the p-chase benchmark from step (2) generate a 2D array of results: one dimension being the different array sizes, the other being the measured time (latency) for each index of the given size. \footnote{We only store first N results of each size -- we do not need to store everything, as the patterns are identical throughout the entire array.}
MT4G reduces the number of dimensions by calculating one reference value from each size vector, implemented as a geometrical mapping technique, introduced in~\cite{HighDimensionCPGGeometricMapping}: 
\begin{equation}
S_i = \sqrt{ \sum_{j=0}^{N} \left( r_{ij} - \min(r) \right)^2 }
\end{equation}
where $S_i$ is the reduced value for index i, $N$ is the number of time measurements in each p-chase size, $r_{ij}$ is the j-th latency, and $min(r)$ is the global minimum latency.
For illustration, Fig.~\ref{fig:raw-data} plots this reduction value (in violet) for different cache types and GPUs.

Every index of the data series $S$ is considered a potential change point, on which the K-S test is applied.
The K-S test denies the null hypothesis when reaching the index of the actual change point, signalling a change point was found, thus identifying the cache size.
The significance level (used to identify the change point) is also reported to provide a confidence metric to the users.

Some approaches, such as the one from~\cite{truong2020selective}, shortlist the change points to evaluate, allowing for a reduction in computation.
Due to our limited dataset size and thus negligible runtime , we omit this step in MT4G.

\subsubsection{Targeting Different Memory Elements}

The size benchmarks target five different memory elements on NVIDIA GPUs and two on AMD GPUs (the remaining ones are retrieved via API).
To ensure targeting the right memory space, the benchmarks are implemented in inline assembly or using intrinsics:
NVIDIA's L1 cache is targeted through global or local loads\footnote{Usually, user code is translated into global loads, while local loads tends to be internal or events, such as register spilling.} --- MT4G uses \texttt{ld.global.ca.u32}: \texttt{.u32} to load a 32-bit unsigned integer, \texttt{.ca} to target all cache levels, and \texttt{.global} to target global memory~\cite{nvidia2025ptx}. 
The texture cache is reached with HIP texture intrinsic instruction \texttt{tex1Dfetch<uint32\_t>(tex, i)}, whose data pass through the texture cache.
Similarly, a readonly cache is addressed with the \texttt{\_\_ldg(\&arr[i])}, where \texttt{arr} is of type \texttt{const uint32\_t*}.
Constant L1 cache is accessed through \texttt{ld.const.u32} instructions: loading a 32-bit unsigned integer from the constant memory, similarly to the Constant L1.5 cache, where the data from the C-L1 cache has to be evicted during the warm-up\footnote{Constant cache array size is limited to \SI{64}{\kibi\byte}, which is the maximal tested size.~\cite{nvidia2025ptx}}.

On AMD GPUs, the sL1 cache is addressed with \texttt{s\_load\_dword} instructions, loading a 32-bit unsigned integer to the scalar general-purpose registers (SGPR) through the sL1 cache.
Similarly, the vector L1 cache is reached with the \texttt{flat\_load\_dword} instruction.~\cite{amd2022mi200}

\subsection{Load Latency benchmarks}

The load latency benchmarks measure the latency of fetching a piece of data from the target memory element using the p-chase pattern, as presented in Section~\ref{sec:p-chase}, however, only with one fixed array size.
Since MT4G can avoid all lower-level caches (if any, with \texttt{.cg} or the GLC bit), it is possible to use a small array of fixed size; MT4G uses size of 256 * Fetch Granularity.

Listings~\ref{lst:p-chase-ptx} and~\ref{lst:p-chase-amdgcn} indicate that the p-chase results already contain the measurement of interest: the latency in clock cycles.
MT4G provides the average as a main result, and a set of statistical values, such as p50, p95, or standard deviation.
It employs assembly instructions and intrinsics to target different memory elements, analogous to Sec.~\ref{sec:size-benchmarks}.
Missing in Sec.~\ref{sec:size-benchmarks}, NVIDIA L2 cache is addressed through \texttt{ld.global.cg.u32}, where \texttt{.cg} bypasses the L1 cache~\cite{nvidia2025ptx}, and Shared Memory is addressed with the \texttt{\_\_shared\_\_} directive.
AMD L2 cache is benchmarked with the same \texttt{flat\_load\_dword} instruction, however, with GLC/sc0 bit set to 1, bypassing the L1 cache.~\cite{amd2022mi200}
LDS uses the \texttt{\_\_shared\_\_} directive.

\subsection{Fetch Granularity benchmarks} \label{sec:fetch-granularity}

A cache line is composed of one or more sectors.
Fetch granularity describes the amount of data fetched to the target cache from higher-level memory.
It can be one or multiple sectors; only a fraction of or the entire cache line.
Only fetching a fraction of a new cache line can, therefore, lead to a subsequent cache miss on the same cache line.
The importance of fetch granularity in data access performance makes it a target for our MT4G microbenchmarking.

On NVIDIA GPUs, one sector is usually \SI{32}{\byte}\footnote{However, there is not enough evidence available that this is a fixed rule.} but, for instance, the V100 GPU's default transaction size is two sectors = \SI{64}{\byte}.
Newer NVIDIA GPU L2 caches have configurable fetch granularity (through the \texttt{cudaDeviceSetLimit} call)~\cite{krashinsky2020optimizing}.

The microbenchmark runs multiple instances of p-chase with strides starting from \SI{4}{\byte}, increasing by \SI{4}{\byte} in each step.\footnote{We assume that the fetch granularity is a multiple of four.}.
MT4G evaluates the results after each run: if both loads and misses were present in the p-chase, some hits must have targeted previously-fetched segments, hence the stride is smaller than the fetch granularity.
Once there are only misses in the p-chase, each element is fetched in a separate transaction, and thus MT4G found the Fetch Granularity of the cache.

\subsection{Cache Line Size benchmarks} \label{sec:cache-line-size}
The size benchmark (Sec.~\ref{sec:size-benchmarks}) works on the following premise: once the p-chase array is larger than the cache size, it will evict cache lines, assuming that the step size is smaller than the cache line size.
Therefore, if we keep increasing the step size, we will see fewer misses beyond the cache size boundary, as not all cache lines are touched (as if the cache was larger).

Since MT4G knows the cache size, it can use this approach and analyze the latencies in p-chase array sizes around and above the cache size, for different step sizes.
It starts with 1/2 of the fetch granularity as a reference (pivot), which surely fits into the cache line, and gradually increases with step size of $fetch\_granularity/2$.
Due to the role of fetch granularity, which also has an impact on misses and aliasing, results of some caches could be interpreted as inconclusive by K-S; hence, we add heuristics to amplify the differences and make the CPD more robust and reliable.

First, we use a pivot and a MAX values that are surely below and above the cache line size, respectively.
We use the pivot as a reference for reduction: the score depends on the similarity to the pivot results, where the larger the array size, the higher the weight the datapoint represents.
Finding the first step size that is more similar to MAX than pivot allows us to adapt all weights above this point to handle aliased outliers.
Finally, we also assume that the cache line size is a power of two.
There is a resulting reduced and heuristically-amplified array at the end, where MT4G searches for the most significant weight increase.

\subsection{Amount benchmarks}\label{sec:amount-bench}

While benchmarks from Section~\ref{sec:size-benchmarks} identify the size of one segment, the Amount benchmarks determine how many of such segments there are per SM/CU or the entire GPU (depending on the memory segment's scope).
It addresses cases like multiple isolated L1 caches per SM, each available to a part of the SM's compute resources.

The workflow uses \textbf{two synchronized cores in one SM/CU}:

\begin{enumerate}
    \item Core A: p-chase warm-up to populate the cache
    \item Core B: p-chase warm-up to populate the cache
    \item Core A: p-chase run to observe hits or misses
\end{enumerate}

\begin{figure}[t]
    \centering
    \includegraphics[width=1.02\linewidth]{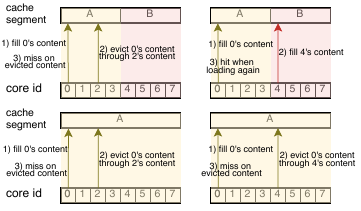}
    \caption{The core of the Amount benchmarks: the two cores evict each others' data if they fetch from the same segment; otherwise not. At the top is a case with 2 segments, at the bottom with one.}
    \label{fig:amount-bench}
\end{figure}

MT4G repeats this approach, as presented in Fig.~\ref{fig:amount-bench}, pinning Core A to Index 0 in the SM/CU, while Core B's index starts with $1$ and doubles this in every repetition, until exceeding the number of cores in the SM/CU.
The p-chase array size is set close to the cache size to ensure potential cache evictions.
If the two cores use the same cache segment, Step (3) will be a cache miss.
If there is only one segment, MT4G always observes a miss (bottom scenario).
If Core B uses a different segment, Core A will hit again in Step (3) (top right scenario), discovering another segment.
In that case, the number of segments MT4G reports is $ NumCoresPerSM / CoreBIndex $.
\subsubsection{L2 segment size}

The L2 cache is a special case: APIs report the total size, while segmentation may limit the size available to SMs/CUs\footnote{For instance, the reported \SI{40}{\mega\byte} L2 cache on NVIDIA A100 is in reality formed by two \SI{20}{\mega\byte} segments~\cite{choquette20213}. Similarly, AMD CDNA3 assumes one L2 cache per XCD~\cite{amd2025gpu}.}.
This flips the previous question: how many segments share the total size reported by the API?

On NVIDIA, the size benchmark is executed to find the size of one L2 segment.
Since the number of segments is an integer, MT4G aligns the segment size to the nearest integer fraction of the API-provided L2 size\footnote{The distance of the benchmarked size from the nearest fraction presents the confidence result of this measurement.}.
On AMD, MT4G assumes one L2 cache per XCD; using the API-provided XCD count.

\subsection{Physical Sharing benchmarks on NVIDIA}
NVIDIA's logical memory spaces (global, texture, constant, ...) may have a dedicated cache hierarchy or share the same physical caches (e.g., \SI{32}{\kilo\byte} L1 and \SI{32}{\kilo\byte}. The texture cache may be one \SI{32}{\kilo\byte} cache).
This distinction is an important topological information.

The Physical Sharing benchmarks are conceptually similar to the Amount benchmarks (Section~\ref{sec:amount-bench}), but they run sequentially on one core only, and target different memory spaces instead:

\begin{enumerate}
    \item P-chase warm-up to populate cache A
    \item P-chase warm-up to populate cache B
    \item P-chase run on cache A to observe hits or misses
\end{enumerate}

If MT4G sees misses in Step (3), the arrays evict each other because they are on the same physical cache, as in the bottom part of Fig.~\ref{fig:amount-bench} if only one core was used, while in the top right of the Figure, the two caches do not interfere with each other.

\subsection{Physical Sharing benchmarks on AMD}
Disregarding the instruction caches, there are only scalar and vector caches available on AMD GPUs, making the NVIDIA approach not applicable.
The Physical sharing benchmark on AMD targets sharing sL1d caches amongst multiple CUs. 
As noted in~\cite{amd2025gpu}, the sL1d cache is shared amongst 2 or 3 CUs, depending on the model.
In reality, the neighboring CUs may be inactive\footnote{For instance, the MI210 GPU has 104 CUs with physical IDs ranging from 0 to 127, indicating the die has 128 CUs with only 104 of those activated.}, resulting in one CU having exclusive access to the entire sL1d cache, compared to others sharing it with 1-2 other CUs.

This benchmark follows the workflow of the Amount benchmark from Section~\ref{sec:amount-bench}, however, it runs with a different configuration:
there are two cores synchronized to work together, but each core is located in a different thread block, scheduled on a specific CU ID.
Respecting this configuration, and ensuring proper synchronization, the benchmark follows the three-step workflow from Section~\ref{sec:amount-bench}.
If the two tested cores (running on two different CUs) evict each others' data, they share the sL1d, otherwise not.
MT4G makes no assumptions about the CU hardware layout and tests all CU pairs.

The output tells the user not only what the sL1d cache size for a specific CU id is, but also tells which CU ids share sL1d, offering two optimization scenarios:
\begin{itemize}
    \item Increase performance when communicating over the sL1d cache across multiple CUs by scheduling the workload on these specific CUs.
    \item Double or triple the available sL1d size by allocating on the CUs whose counterpart is disabled.
\end{itemize}

\subsection{Bandwidth Benchmarks} \label{sec:bw-bench}
The bandwidth benchmarks work differently from the rest of the benchmarks -- neither are they based on the p-chase pattern, nor do they run on one thread only.
The benchmark is inspired by the stream data access pattern, known by similar approaches~\cite{mccalpin2006stream,jia2019dissecting}.
It executes frequent load (or store) instructions, each accessing different memory locations, and runs with the maximum number of threads per block (which MT4G reports) to ensure that enough loads are in the pipeline.
It is scheduled with a specific number of blocks (our experiments heuristically found that $num\_SMs * max\_blocks\_per\_SM$ blocks result in the highest bandwidth).

MT4G uses vector loads, fetching \SI{128}{\bit} of data in one instruction (\texttt{ld.global.v4.u32} on NVIDIA, loading 4 x 32-bit unsigned integers from the global memory, and \texttt{flat\_load\_dwordx4} on AMD with the same effect (4 dwords = \SI{128}{\bit}).
Coalesced memory access pattern minimizes the amount of memory transactions, thus maximizing throughput~\cite{fauzia2015characterizing}.
The benchmark, which runs multiple threads and blocks in parallel, uses \texttt{hipDeviceSynchronize()} to synchronize the start and finish of the run to ensure concurrent execution and thus correct reporting.
\texttt{hipEventRecord(start/end)} measures the kernel execution time; the resulting bandwidth is the total amount of data loaded over the kernel execution time.

\section{Validation of the Results} \label{sec:validation} 

\newcommand{\colorsquare}[1]{\textcolor{#1}{\rule{1ex}{1ex}}}
\newcommand{\na}[1]{\text{\textit{\textcolor{gray}{n/a}}}}
\newcommand{\textbful}[1]{\textbf{\underline{#1}}}

\begin{table*}[t]
    \centering
    \begin{tabular}{| c || c | c | c | c | c |}
    \hline
    \textbf{GPU Name} & Vendor & Microarch. & GPU & CPU & OS\&Software$^*$ \\
    \hline
    \textbf{P6000} & NVIDIA & Pascal & Quadro P6000 & {\footnotesize Intel(R) Xeon(R) Gold 6238} & {\footnotesize Ubuntu 22.04; 6.3; 12.8; 570.158.01} \\
    \textbf{V100} & NVIDIA & Volta & V100 16GB & {\footnotesize Intel(R) Xeon(R) Gold 6238} & {\footnotesize Ubuntu 22.04; 6.3; 12.8; 570.158.01}\\
    \textbf{T1000} & NVIDIA & Turing & T1000 & {\footnotesize Intel(R) Xeon(R) Silver 4116} & {\footnotesize Ubuntu 24.04; 6.1.2; 12.9; 570.133.20}\\
    \textbf{RTX2080} & NVIDIA & Turing & GeForce RTX 2080 Ti & {\footnotesize AMD Ryzen Threadripper 2990WX} & {\footnotesize Ubuntu 24.04; 6.1.2; 12.9; 570.158.01}\\
    \textbf{A100} & NVIDIA & Ampere & A100 & {\footnotesize AMD Ryzen Threadripper PRO 3955WX} & {\footnotesize Ubuntu 24.04; 6.3.0; 12.9; 570.158.01 }\\
    \textbf{H100-80} & NVIDIA & Hopper & H100 80GB HBM3 & {\footnotesize AMD EPYC 9374F 32-Core Processor} & {\footnotesize Rocky 9.1; 6.4; 12.9; 535.54.03}\\
    \textbf{H100-96} & NVIDIA & Hopper & H100 96GB HBM3 & {\footnotesize AMD EPYC 9374F 32-Core} & {\footnotesize Ubuntu 24.04; 6.4; 12.9; 570.172.08 }\\
    \hline
    \textbf{MI100} & AMD & CDNA & Instinct MI100 & {\footnotesize AMD EPYC 7742 64-Core Processor }& {\footnotesize SLES15; 6.4; 6.10.5 }\\
    \textbf{MI210} & AMD & CDNA2 & Instinct MI210 & {\footnotesize AMD EPYC 7773X 64-Core Processor}& {\footnotesize SLES15; 6.3.3; 6.10.5} \\
    \textbf{MI300X} & AMD & CDNA3 & Instinct MI300X VF & {\footnotesize Intel(R) Xeon(R) Platinum 8568Y+}& {\footnotesize Ubuntu 24.04; 6.4; 6.12.12}\\
    \hline
    \end{tabular}
    \raggedright{\footnotesize{$^*$: NVIDIA: <OS, hipcc version, nvcc version, driver version>; AMD: <OS, hipcc version, ROCk version>.}}\\
    \caption{Specifications of servers and GPUs where we validate MT4G results.}
    \label{tab:validated-on}
\end{table*}

We validate the MT4G's microbenchmark results across 10 GPUs (7 NVIDIA, 3 AMD GPUs; cf.~Table~\ref{tab:validated-on}) to showcase the universality, accuracy, and reliability of the auto-evaluated benchmarks.
The results are composed of discrete attributes (cache line size, fetch granularity, amount, physical layout), where any error results in a wrong result, and continuous ones (size, load latency, bandwidth), where minor errors are an inevitable measurement artifact.

Table~\ref{tab:results} presents MT4G results for one recent GPU of each vendor: NVIDIA H100 and AMD MI210 GPUs, against reference sources taken from official documentation, peer-reviewed papers, other sources, or previous microarchitectures\footnote{Comparing attributes like cache sizes and latency brings no additional value, as these change with every microarchitecture; while microarchitectural details, such as cache line size may (or may not) stay over multiple generations.}, in this order of availability.
\textbf{Bold} results denote information unavailable for the presented GPU, while \underline{\textbf{bold underlined}} reveal new information. 

The discrete attributes \textbf{always match the references, and often reveal new information}.
Cache line sizes and fetch granularities are two examples worth mentioning: MT4G closes the knowledge gap by revealing these previously unavailable attributes, which are critical in determining and optimizing cache behavior for many applications.
On the AMD side, identifying the ids of CU that share one sL1d cache presents a new opportunity for specific kernels that can harness the sharing or isolation properties.

From the continuous attributes, cache size on lower-level NVIDIA caches slightly deviates from the reference, which, however, describes a total L1+Shared memory size, which can be split during runtime\footnote{\texttt{cudaDeviceSetCacheConfig} options PreferShared/L1/Equal; The config, an MT4G CLI parameter, can be configured to measure any of these; by default and in this Section PreferL1 was used.}.
Therefore, MT4G provides an additional detail: \textbf{the true L1 size}.
The exception is the Constant L1.5 cache, measuring up to \SI{64}{\kibi\byte} (cf.~Sec.~\ref{sec:scope}); however, a confidence of 0 makes it clear that it is not the actual size.
Consequently, the CL1.5 cache line size, which takes the cache size as input, is not computed.
On AMD, the results match or are very close to official reference numbers.

The official bandwidth information is not further specified, presumably providing theoretical values, while MT4G provides achieved throughput -- a metric that is easier to grasp.
MT4G bandwidth benchmark achieves \SI{20}{\percent} lower bandwidth than reports~\cite{NvidiaH100_FunnyL2_chipsandcheese} on NVIDIA L2, indicating that more optimal thread and block dimensions exist; however, 1) the provided values are sufficient for a bandwidth estimate, 2) MT4G is not tuned to specific hardware, and 3) MT4G also provides the read bandwidth.
The reported latencies stay very close to the results used in other studies.
As no reference is available, we cannot argue which results are "better".
As the primary purpose of providing the latencies is comparing different caches and estimating the benefit or penalty of hitting or missing a cache, both the other available and our results clearly answer these questions.
Some of the latencies on these devices were previously unavailable, making us the first to extract them.
Since AMD MI210 features no L3 cache, no results are presented.

Evaluating the remaining eight machines, we mainly observe identical behavior: the discrete attributes provide the correct information matching with available sources, and present new findings where not. The continuous attributes present some deviations from the official sources (cache size, if not obtained by an API), however, the deviations from the available sources match the one presented above.

\begin{table*}[t]
    \centering
    \begin{tabular}{| c | c || c | c | c | c | c | c | c |}
    \hline
    \textbf{Component} & & \textbf{Size} & \textbf{\makecell{Load \\ Latency}} & \textbf{\makecell{R \& W \\ Bandwidth}} & \textbf{\makecell{Cache \\ Line Size}} & \textbf{\makecell{Fetch \\ Granularity}} & \textbf{\makecell{\# per SM/CU \\ or GPU}} & \textbf{\makecell{Physically \\ Shared With}} \\
    \hline
    \hline
\multicolumn{9}{|c|}{\raisebox{-0.5ex}{\LARGE \textbf{NVIDIA H100-80 SXM5}}} \\
    \hline
    \multirow{2}{*}{\textbf{L1}} & {\small Ref} & \colorsquare{green}~256KB~\cite{nvidia_h100_2022} & \colorsquare{cyan}~$30\text{-}40$~\cite{jarmusch2025dissectingnvidiablackwellarchitecture} & \multirow{2}{*}{\na} & \colorsquare{Orchid3}~32B~\cite{jia2018dissecting}  & \colorsquare{Orchid3}~32B~\cite{jia2018dissecting} & \colorsquare{green}~1~\cite{nvidia_h100_2022}  & {\small \colorsquare{Orchid3}~RO,TX,L1~\cite{GPUPascalSMArch}$^1$} \\
     & {\small MT4G} & 238KiB & 38 & & \textbf{128B} & \textbf{32B} & 1 & {\small \textbf{RO,TX,L1}} \\
    \hline

	\multirow{2}{*}{\textbf{L2}} & {\small Ref} & \colorsquare{green}~50MB~\cite{nvidia_h100_2022} & \colorsquare{cyan}~273~\cite{jarmusch2025dissectingnvidiablackwellarchitecture} & \colorsquare{DarkGoldenrod3}~5.56TB/s~\cite{NvidiaH100_FunnyL2_chipsandcheese} & \colorsquare{Orchid3}~64B~\cite{jia2018dissecting} & ? & \colorsquare{green}~2~\cite{nvidia_h100_2022}  & {\small \multirow{2}{*}{\na}} \\
     & {\small MT4G} & \colorsquare{lightgray}~50MB & 220 & 4.4/3.4 TiB/s & \textbf{128B}  & \textbful{32B} & 2 &  \\
    \hline

	\multirow{2}{*}{\textbf{Texture}} & {\small Ref} & \colorsquare{green}~256KB~\cite{nvidia_h100_2022} &  ?  & \multirow{2}{*}{\na} & ?  & ? & \colorsquare{Orchid3}~1~\cite{GPUPascalSMArch} & {\small \colorsquare{Orchid3}~RO,TX,L1~\cite{GPUPascalSMArch}$^1$} \\
     & {\small MT4G} & 238KiB & \textbful{39} &  & \textbful{128} & \textbful{32} & \textbf{1} & {\small \textbf{RO,TX,L1}} \\
    \hline

	\multirow{2}{*}{\textbf{Readonly}} & {\small Ref} & \colorsquare{green}~256KB~\cite{nvidia_h100_2022} & ? & \multirow{2}{*}{\na} & ? & ? & \colorsquare{Orchid3}~1~\cite{GPUPascalSMArch} & {\small \colorsquare{Orchid3}~RO,TX,L1~\cite{GPUPascalSMArch}$^1$} \\
     & {\small MT4G} & 238KiB & \textbful{35} &  & \textbful{128} & \textbful{32} & \textbf{1} & {\small \textbf{RO,TX,L1}} \\
    \hline

	\multirow{2}{*}{\textbf{Const L1}} & {\small Ref} & ? & ? & \multirow{2}{*}{\na} & \colorsquare{Orchid3}~64B~\cite{jia2018dissecting} & ? & \colorsquare{Orchid3}~?~\cite{jia2018dissecting} & {\small ?} \\
     & {\small MT4G} & \textbf{2KiB} & \textbf{21} & & \textbf{64B} & \textbful{64B} & \textbf{1} & {\small \textbf{no}}\\
    \hline

	\multirow{2}{*}{\textbf{Const L1.5}} & {\small Ref} & ? & ? & \multirow{2}{*}{\na} & \multirow{2}{*}{\na} & ? & \multirow{2}{*}{\na} & {\small \multirow{2}{*}{\na}} \\
     & {\small MT4G} & \textbf{>64KiB} & \textbf{105} &  & & \textbful{256B} & &  \\
    \hline

	\multirow{2}{*}{\textbf{Shared Mem}} & {\small Ref} & \colorsquare{green}~228KB~\cite{nvidia_h100_2022} & ? & \multirow{2}{*}{\na} & \multirow{2}{*}{\na} & \multirow{2}{*}{\na} & \multirow{2}{*}{\na} & {\small \multirow{2}{*}{\na}} \\
     & {\small MT4G} & \colorsquare{lightgray}~228KiB & \textbf{30} & & & & & \\
    \hline

	\multirow{2}{*}{\textbf{Device Mem}} & {\small Ref} & \colorsquare{green}~80GB~\cite{nvidia_h100_2022} & \colorsquare{cyan}~658~\cite{jarmusch2025dissectingnvidiablackwellarchitecture} & \colorsquare{green}~3.35 TB/s~\cite{nvidia_h100_2022_datasheet} & \multirow{2}{*}{\na} & \multirow{2}{*}{\na} & \multirow{2}{*}{\na} & {\small \multirow{2}{*}{\na}} \\
     & {\small MT4G}& \colorsquare{lightgray}~80GB & 843 & 2.5/2.7 TiB/s &  &   &   &   \\

    \hline
    \hline
    \multicolumn{9}{| c |}{\raisebox{-0.5ex}{\LARGE \textbf{AMD Instinct MI210}}} \\
    \hline

    \multirow{2}{*}{\textbf{vL1}} & {\small Ref} & \colorsquare{green}~16KiB~\cite{ amd2025gpu} & \colorsquare{DarkGoldenrod3}~145~\cite{ADMInstinctMI210_chipsandcheese} & \multirow{2}{*}{\na} & \colorsquare{green}~64B~\cite{amdrocprofiler} & ? & \colorsquare{green}~1~\cite{ amd2025gpu} & {\small \multirow{2}{*}{\na}} \\
     & {\small MT4G} & 16KiB & 125 & & 64B & \textbful{64B} & 1  & \\
	\hline

    \multirow{2}{*}{\textbf{sL1d}} & {\small Ref} & \colorsquare{green}~16KiB~\cite{ amd2025gpu} & \colorsquare{DarkGoldenrod3}~64~\cite{ADMInstinctMI210_chipsandcheese} & \multirow{2}{*}{\na} & ? & ? &  & {\small \colorsquare{green}~\# CUs~\cite{ amd2025gpu}} \\
     & {\small MT4G} & 15.5KiB & 50 & & \textbful{64B} & \textbful{64B} & \multirow{-2}{*}{\na} & {\small \textbful{CU id}}\\
	\hline
    
    \multirow{2}{*}{\textbf{L2}} & {\small Ref} & \colorsquare{green}~8MB~\cite{ amd2025gpu} &  ? & \colorsquare{DarkGoldenrod3}~3.7 TB/s~\cite{ADMInstinctMI210_chipsandcheese} & \colorsquare{green}~128B~\cite{amdrocprofiler} & ? & \colorsquare{green}~1~\cite{amd2021cdna2whitepaper} & {\small \multirow{2}{*}{\na}} \\
     & {\small MT4G} & \colorsquare{lightgray}~8MB & \textbful{310} & 4.19/2.4 TiB/s & \colorsquare{lightgray}~128B & \textbful{64B} & \colorsquare{lightgray}~1 & \\
    \hline

	\multirow{2}{*}{\textbf{L3}} & {\small Ref} & & & & & \multirow{2}{*}{\na} & & {\small \multirow{2}{*}{\na}} \\
     & {\small MT4G} & \multirow{-2}{*}{~\na} & \multirow{-2}{*}{\na} & \multirow{-2}{*}{\na} & \multirow{-2}{*}{\na} & & \multirow{-2}{*}{\na} &   \\
    \hline

	\multirow{2}{*}{\textbf{LDS}} & {\small Ref} & \colorsquare{green}~64KiB~\cite{ amd2025gpu} & \colorsquare{DarkGoldenrod3}~61~\cite{ADMInstinctMI210_chipsandcheese} & \multirow{2}{*}{\na} & \multirow{2}{*}{\na} & \multirow{2}{*}{\na} & \multirow{2}{*}{\na} & {\small \multirow{2}{*}{\na}} \\
     & {\small MT4G} & \colorsquare{lightgray}~64KiB & 55 &   &   &   &   &  \\
    \hline

	\multirow{2}{*}{\textbf{Device Mem}} & {\small Ref} & \colorsquare{green}~64GB~\cite{ amd2025gpu} & ? & \colorsquare{green}~1.6 TB/s~\cite{amd2021cdna2whitepaper} & \multirow{2}{*}{\na} & \multirow{2}{*}{\na} & \multirow{2}{*}{\na} & {\small \multirow{2}{*}{\na}} \\
     & {\small MT4G} & \colorsquare{lightgray}~64GB & \textbful{748} & 1.0/0.9 TiB/s &  &  &  & \\

    \hline
    \end{tabular}

    \raggedright{
    $^1$ = Since Pascal: L1, Texture and Read-Only Cache are unified, no longer explicitly stated in official docs of current architectures\\
    \colorsquare{green} = Official references ;
    \colorsquare{cyan} = Peer-reviewed references ;
    \colorsquare{Orchid3} = Peer-reviewed references regarding other GPUs ;
    \colorsquare{DarkGoldenrod3} = Other references \\
    \colorsquare{LightGray} = Retrieved by MT4G via API}
    \caption{Comparison of available information about memory components against tool results.}
    \label{tab:results}
\end{table*}

Apart from the Constant L1.5  and line sizes, three of the benchmarks did not provide a final result (no result or zero confidence in the report, not a wrong result) on one GPU:
\begin{enumerate}
    \item MI300X CU id: Due to our virtualized environment, access to the GPU, CU id sharing benchmarks could not be executed.
    \item P6000 L1 Amount: On the Pascal GPU, MT4G was unable to identify the amount of L1 caches per SM. 
    The benchmark works on other Pascal caches as well as L1 caches on other machines, making this a unique combination.
    MT4G was unable to schedule a thread on warp 3 (of 4), making it impossible to perform the benchmark as planned (cf. Sec.~\ref{sec:amount-bench}).
    \item P6000 L1 Shared with Constant L1: the Shared benchmark sometimes incorrectly indicates L1 and Constant L1 cache sharing on Pascal; being likely a related issue to the L1 amount on the same GPU architecture. We will investigate this issue and fix it in a future release.
\end{enumerate}

The provided evaluation shows that MT4G presents reliable results across vendors and microarchitectures.
The information, otherwise uneasy to find or not available at all, highlights MT4G's utility as a vital tool for understanding GPU topologies.

\subsection{Run Times}

The run times vary significantly between vendors (\SIrange{6}{14}{\minute} on NVIDIA and ca. \SI{1}{\minute} on AMD).
This is due to the different number of benchmarks executed on the machines (35 vs. 15), where especially the L2 benchmarks have a significant impact on the total run time (\SI{4.5}{\minute} out of a total of \SI{12.25}{\minute} on A100), as the benchmarks must repeatedly fill the large L2 cache and beyond.
Since we see MT4G being executed infrequently, ca. \SI{15}{\minute} and \SI{1}{\minute} runtime is acceptable.
Moreover, users can limit MT4G execution to a specific memory segment through the CLI\footnote{While the entire run takes over \SI{12}{\minute} on A100, L1-only run takes just over \SI{1}{\min}.} or configure the measurements more coarsely and thus significantly reduce the run time.

\section{Utilizing MT4G}

Beyond presenting the microbenchmarks (Section~\ref{sec:benchmarks}) and validating their precision (Section~\ref{sec:validation}), this section illustrates the \textbf{integration of MT4G into three different scenarios}: GPU performance modeling, performance optimizations, and dynamic topology management.
These use-cases demonstrate the need for a tool, such as MT4G, and its ability to provide automated, vendor-agnostic, and reliable information about the GPU topology.

\subsection{GPU Performance Modeling}
Performance modeling helps to understand hardware-software interactions and provides quantitative estimates for the expected performance of an application or set of applications on a given system.
In this use case, we focus on utilizing GPU warp parallelism-based performance model~\cite{hong2009analytical} for understanding the performance of GPU applications, along with key resource bottlenecks. In the proposed model, there are two key categories of model parameters: (1) Application-specific parameters, and (2) GPU-specific parameters. For effectively utilizing this model to estimate the performance of a given application, it is important to obtain accurate values for the model parameters. 
For GPU-specific parameters, one needs various hardware characteristics, such as compute resource information including maximum registers per SM/CU, maximum threads per SM/CU, maximum active warps per SM/CU and warp size, and memory resource information including bandwidth and latency across GPU memory hierarchy (L1, L2, and DRAM). Some of these parameters could be obtained from vendor-specific GPU architecture documentation, but not all parameters are available in the documentation, and some of them may not be accurate.
MT4G provides a reliable and vendor-agnostic way to obtain these GPU-specific parameters, which can be used to build the performance model. Hence, we integrate MT4G into the performance modeling workflow to obtain the GPU-specific parameters.

In the given model formulation, we consider compute warp parallelism (CWP) and memory warp parallelism (MWP) as the two key performance indicators, where CWP (Equation~\ref{eq:cwp}) is defined as the number of warps that can execute concurrently while a warp waits for memory transfers, and MWP (Equation~\ref{eq:mwp}) is defined as the number of warps that can access the memory subsystem concurrently.

\begin{equation}\label{eq:cwp_mem_wait}\nonumber
    CWP'=\frac{mem\_cycles + comp\_cycles}{comp\_cycles}
\end{equation}
\begin{equation}\label{eq:mwp_latency}\nonumber
    MWP'=\frac{mem\_latency}{mem\_delay}
\end{equation}
\begin{equation}\label{eq:mwp_bandwidth}\nonumber
\hspace{-3pt}    MWP''=\frac{mem\_bandwidth}{\frac{mem\_freq \times load\_per\_warp}{mem\_latency} \times  \#act\_warps\_per\_SM}
\end{equation}
\begin{equation}\label{eq:cwp}
CWP = min(CWP', \#act\_warps\_per\_SM)
\end{equation}
\begin{equation}\label{eq:mwp}
    MWP = min(MWP',MWP'',\#act\_warps\_per\_SM)
\end{equation}

While the parameters required for CWP depend on application characteristics, those for MWP are primarily GPU-specific and can be obtained using MT4G –- namely, $mem\_latency$, $mem\_bandwidth$, and $mem\_freq$. The original model formulation~\cite{hong2009analytical} focuses solely on main memory transfers, but it can be extended to include the L1/L2 cache, as MT4G provides these parameters across the memory hierarchy, including L1, L2, and DRAM. By combining MT4G with application profiling using NVIDIA Nsight Compute~\cite{nsightcompute} or AMD ROCProfiler~\cite{amdrocprofiler}, we can construct an accurate performance model for an application on a given GPU.

Using these parameters, we can analyze resource bottlenecks: if CWP exceeds MWP, the application is memory-bound; otherwise, it is compute-bound. Ultimately, the mentioned formulations for CWP and MWP shall be used to estimate the elapsed GPU cycles for a given application. Further details on the performance model and the equations for calculating GPU cycles are available in the work by Hong et al.~\cite{hong2009analytical}. Lastly, these parameters obtained via MT4G can also support the development of accurate and vendor-agnostic models using other methods, such as the Roofline model~\cite{williams2009roofline}.

\subsection{GPUscout GUI}

After performance modeling, we present the integration of MT4G in the workflow of GPUscout~\cite{sen2023gpuscout} --- a performance analysis tool, which detects memory-related bottlenecks in the kernels, and gives recommendations to users, backed by NCU counters.
These recommendations are closely tied to the GPU topology: for instance, register spilling is tied to the number of cores and registers per SM, or the L1 hit rate is tied to the L1 size.

Struckenberger's Graphical User Interface (GUI) extension of GPUscout~\cite{stuckenberger2025} facilitates and deepens the understanding of the bottleneck and the machine state, and the options to resolve said bottlenecks.
The machine state is presented through combining the NCU counters with the GPU compute and memory topology.
Next to the GPUscout analysis output, \textbf{GPUscout GUI integrated MT4G into its workflow} to provide the hardware topological context\footnote{GPUscout GUI currently parses the original MT4G CSV output, and will be migrated to JSON in the future.}.

\begin{figure}[t]
    \centering
    \includegraphics[width=1.0\linewidth]{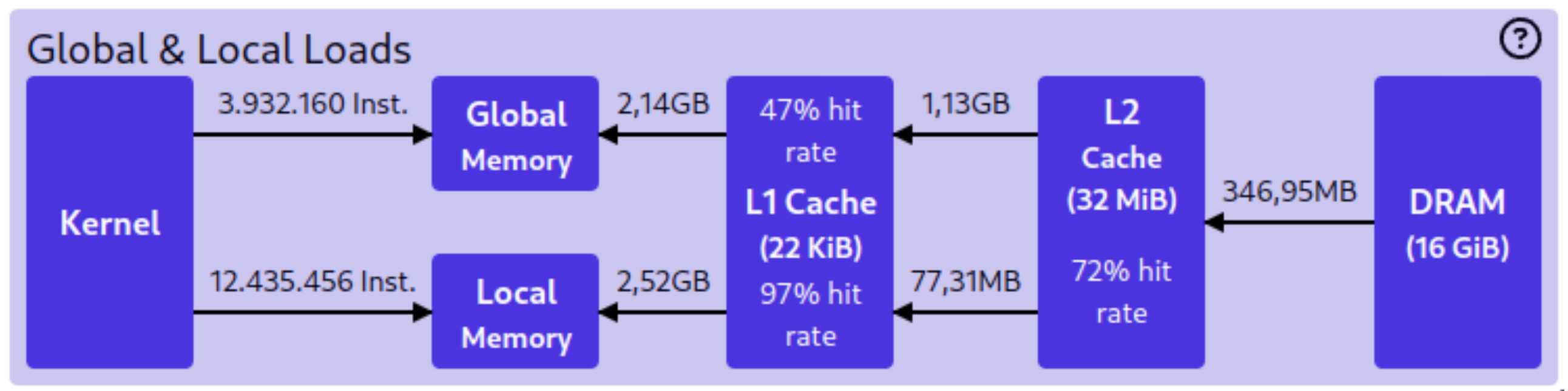}
    \caption{GPUscout-GUI Memory Component visualization, containing memory element sizes.~\cite{stuckenberger2025}.}
    \label{fig:gpuscout}
\end{figure}

The MT4G integration manifests itself in multiple views in the GUI, such as in the Memory Graph component illustrated in Fig.~\ref{fig:gpuscout}.
The GPUscout-provided (via NCU) metrics, such as the traffic between the caches or the hit rate, are complemented with the MT4G information, such as cache sizes.
For instance, knowing the L1 cache size helps in understanding the cache miss rate.
Or, the L1 size is a crucial attribute needed to decide whether and how to redesign block dimensions of a problem to fit in the cache.
Without the MT4G information, users would have to guess these parameters, hoping an arbitrary change improves performance.

This integration enhances the capabilities of GPUscout-GUI, a new visualization interface of an established tool GPUscout, enabling better-informed and more efficient performance tuning on target GPUs.
Future releases of GPUscout (and GPUscout-GUI) covering AMD GPUs could seamlessly integrate the already vendor-agnostic report of MT4G, lowering the future integration burden.

\subsection{Integration with sys-sage for performance estimation}\label{sec:sys-sage}

The sys-sage library~\cite{vanecek2024sys,vanecek2023syssage} manages HPC system topologies, CPUs, QPUs (Quantum Processing Units)~\cite{mishra2025towards}, and \textbf{GPUs -- thanks to MT4G integration}.
Static MT4G context is combined with dynamic GPU configuration queries, retrieving resource isolation settings (MIG -- Multi-instance GPU)~\cite{mig-techreport}.
MIG partitions GPU compute resources, cache, memory, and bandwidth, allowing multiple jobs or even users to utilize the same GPU without interference.

\begin{figure}[t]
    \centering
    \includegraphics[width=1.0\linewidth]{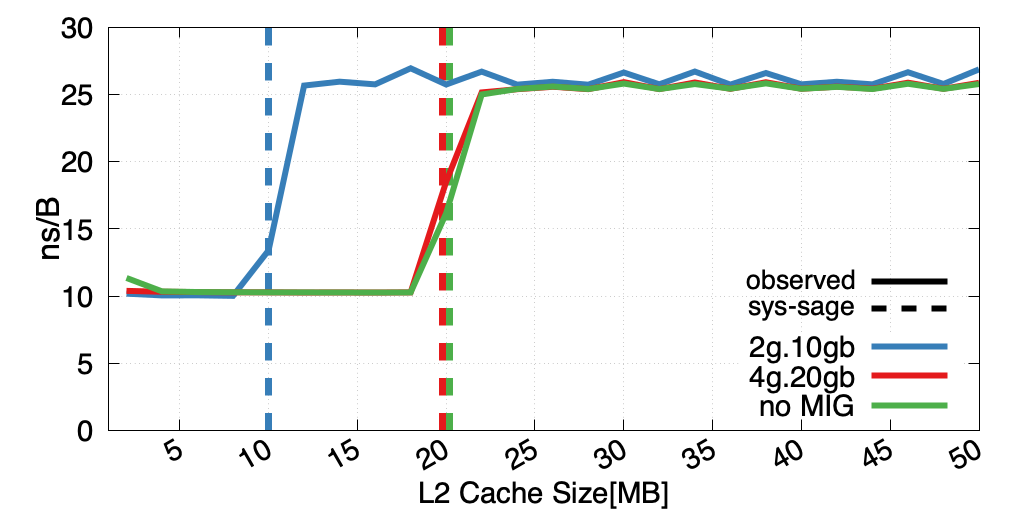}
    \caption{Streaming read throughput over an array with varying sizes on NVIDIA A100 with differnent MIG settings. Vertical lines mark L2 cache size provided by sys-sage~\cite{vanecek2024sys}.}
    \label{fig:sys-sage-mig}
\end{figure}

Fig.~\ref{fig:sys-sage-mig} shows how static MT4G information is combined with dynamic MIG settings (the \texttt{nvml} library) through integration in sys-sage to provide current GPU capabilities.
The y-axis shows performance (\SI[per-mode=repeated-symbol]{}{\nano\second\per\byte}) of the stream algorithm with different vector sizes, running on one NVIDIA A100 GPU core.
Vertical dashed lines denote the L2 cache size reported by sys-sage, combining \textbf{MT4G input} with different MIG settings.
Two observations can be made:

\begin{enumerate}
    \item A steep performance drop beyond the reported L2 cache size validates the value reported by sys-sage and indicates a problem size boundary for higher performance.
    \item No difference exists between the full GPU and the `\texttt{4g.20gb}' MIG setting (4 out of 7 GIs, \SI{20}{\mega\byte} L2 cache, and \SI{20}{\giga\byte} of main memory), as one SM can anyways only access one of the two \SI{20}{\mega\byte} L2 cache segments (cf. Sec.~\ref{sec:amount-bench}).
    Without MT4G's correct L2 Amount information, the performance drop would not match the reported L2 size, hurting the performance of kernels relying on this information.
\end{enumerate}

This scenario highlights performance penalties associated with inefficient GPU resource usage due to poorly chosen problem sizes or domain divisions.
This issue becomes even more complex with dynamic runtime settings, such as NVIDIA MIG, underscoring the importance of the topological details provided by MT4G.
While APIs such as \texttt{nvml} or LIKWID lack sufficient detail, integrating MT4G into sys-sage enables an accurate understanding of dynamic runtime configuration and its effects on topology.

\section{Conclusions and Future Work}

In this work, we introduced MT4G, an open-source, vendor-agnostic tool for auto-discovering GPU compute and memory topologies. 
By leveraging existing APIs and over 50 automated microbenchmarks, MT4G provides comprehensive, reliable reports on GPU topology and attributes, including cache sizes, load latencies and bandwidth, cache line size, fetch granularity, and hardware-topological layouts across NVIDIA and AMD GPUs.
Based on the fine-grained p-chase and robust statistical evaluation methods like the Kolmogorov-Smirnov test, a set of benchmarks is automatically evaluated and provides outlier-resistant results of these key metrics and properties. 

We validated MT4G's benchmarks across 10 GPUs against official and one-off reverse-engineering sources to show the accuracy of the provided information.
Finally, we demonstrated the utility in three scenarios where MT4G was integrated into other tools' workflows to provide needed information: GPU performance modeling toolchain, providing hardware context to performance analysis with GPUscout, and enabling dynamic topology management through integration into sys-sage.
These integrations highlight MT4G's value, enabling optimized performance and resource allocation for HPC and AI workloads across NVIDIA and AMD GPUs.

\subsubsection*{Future Work}

To fill the gaps in AMD L3 cache benchmarking, we will conduct more experiments on the CDNA3 microarchitecture.
Second, we plan to extend the bandwidth benchmarking to low-level caches.
We also plan to incorporate compute capability metrics, such as FLOPS for INT and FP datatypes of different precisions, addressing differences in HPC and AI workloads.
We aim to characterize specialized engines, like tensor cores, to provide detailed insights into their capabilities.
Finally, to ensure MT4G's relevance in the future, we will validate emerging architectures, like NVIDIA Blackwell or AMD CDNA4, and adapt the benchmark suite to address new architectural specifics.



\appendix
\subsection{Artifact Description}
\subsubsection{Overview of Contributions and Artifacts}
\paragraph{Paper's Main Contributions}

\begin{description}
\item[$C_1$] The automated tool MT4G that reports on GPU topologies.
\item[$C_2$] The set of microbenchmarks to reverse-engineer GPU topologies.
\item[$C_3$] Automated evaluation of the benchmarks using the \linebreak \mbox{Kolmogorov-Smirnov} test and other statistical methods.
\end{description}

\paragraph{Computational Artifacts}

\begin{description}
\item[$A_1$] https://doi.org/10.5281/zenodo.16788785
\end{description}

\begin{center}
\begin{tabular}{rll}
\hline
Artifact ID  &  Contributions &  Related \\
             &  Supported     &  Paper Elements \\
\hline
$A_1$   &  $C_1$ & Table 1\\
\hline
$A_1$   &  $C_2$ & Figures 1-3 \\
        &        & Tables 1-2\\
\hline
$A_1$   &  $C_3$ & Figure 2 \\
        &        & Table 3 \\
\hline
\end{tabular}
\end{center}

\subsubsection{Artifact Identification}
\paragraph{Relation To Contributions}

The artifact contains the source code of MT4G, which can be compiled and run.
Moreover, it provides additional information about machines from Table 2, used to generate results presented in Sec.5, Table 3.
Finally, it contains raw results and plots that are a basis for Figure 2.

\paragraph{Expected Results}
In this artifact is the source code of MT4G -- one can build it and run it to analyze an NVIDIA (Pascal or newer) or AMD (CDNA) GPUs, as described in the paper.

Moreover, the artifact contains 1) additional details that were not discussed in the paper Section 5, and 2) it contains the raw results and the JSON output of the other presented GPUs from Table 2.

One will be able to reproduce all results presented in Section 5, provided the user has the access to the same machines with the same configuration.
Evaluating MT4G on the same GPUs with the same compiler, however a different CPU should also lead to comparable results but we would have to conduct more experiments to be sure about that statement.
In general, one will get a report of an arbitrary supported GPU in the detail described in Section 3.
One can then verify the results using the available sources, where available.

\paragraph{Expected Reproduction Time (in Minutes)}

\paragraph*{Artifact Setup}
The time needed for the artifact setup strongly depends on the available software on the machine. Installing all dependencies via spack takes ca 60 min per machine.
Then, MT4G code compiles within 10 minutes.

\paragraph*{Artifact Execution}
Once all the dependencies and MT4G are installed, one complete analysis takes ca 15 min on a single NVIDIA GPU, or ca 2 min on AMD CDNA GPU.
The time may increase if borders need to be widened for some experiments.
Repeating all of them sequentially, should therefore take approximately 2 hours.

\paragraph*{Artifact Analysis}
A substantial amount of time can be spent looking for available reference results, if the GPU differs from the ones presented in Section 5.
Assuming the tests are run on the same hardware, one just needs to compare the output JSON reports, requiring approximately 15 min.

\paragraph{Artifact Setup (incl. Inputs)}

\paragraph*{Hardware}

An NVIDIA GPU, Pascal or newer, Hopper was the newest architecture tested. On AMD, GPU of CDNA 1, 2, or 3 is necessary.
To reproduce our measurements, please refer to Table 2 from the paper to see details on used hardware specifications.

\paragraph*{Software}

\begin{itemize}
    \item HIP SDK with the `hipcc` compiler  \url{https://packages.spack.io/package.html?name=hip} (`hip` for AMD, `hip+cuda` for NVIDIA) please refer to Table 2 for versions (6.1+)
    \item (NVIDIA GPUs) nvcc compiler, \url{https://docs.nvidia.com/cuda/cuda-compiler-driver-nvcc/}, version 12.8 or 12.9
    \item Python 3 with the `matplotlib`, `pandas` and `numpy` packages; most recent version
\end{itemize}

\paragraph*{Datasets / Inputs}

No datasets are necessary as input.

\paragraph*{Installation and Deployment}

Please refer to the Readme for the build information.
In short, MT4G is built with make; the build command is

\begin{lstlisting}
    make GPU_TARGET_ARCH=<sm_XX|gfxXXX>
\end{lstlisting}

where the GPU\_TARGET\_ARCH is the target architecture, e.g, \texttt{sm\_90} for NVIDIA Hopper or \texttt{gfx90a} for AMD CDNA2.
Note: If you execute make in parallel (-j), it may fail during the first build; in that case, please reexecute the same make command again.

\paragraph{Artifact Execution}

There is only one step, which is executing the compiled MT4G binary -- depending on the CLI parameters, it adjusts the run parameters and outputs the report JSON (stdout by default, can be changed).

The runs with outputs available in artifact $A_1$(folder \texttt{results}) were generated executing the following command: 

\begin{lstlisting}
    ./mt4g -g -o -p -j
\end{lstlisting}

$-g$ to generate graphs, such as ones from Fig. 2; $-o$ to store raw timing data; $-p$ to create a .md report; $-j$ to create the JSON output file.

Running \texttt{./mt4g} is sufficient: it prints the JSON to stdout.
All measurements including time were collected with \texttt{./mt4g -q} command -- avoiding the overhead of generating other output than JSON. 

\paragraph{Artifact Analysis (incl. Outputs)}

The main item in the generated output is the JSON report (printed to stdout by default or <GPU\_name>.json to current directory if option $-j$ is enabled).

Using an equivalent hardware, one can verify the results of Section 5 of the paper, or one can refer to artifact $A_1$'s \texttt{results/} folder to compare the JSON outputs directly.

Moreover, if executed with option $-g$, a results folder \\(\texttt{mt4g/results/<GPU\_name>}) is generated with plots such as those presented in Fig. 2. 
They can be used to analyze and evaluate the CPD algorithms used, as well as the benchmark behavior.
Again, they can be compared with results stored in artifact $A_1$'s \texttt{results/} folder.

\bibliographystyle{IEEEtran}
\bibliography{IEEEabrv,sample-base}
\end{document}